\documentclass[12pt]{article}
\usepackage{amsfonts}
\usepackage{amsmath,amssymb,array}
\numberwithin{equation}{section}
\def\be{\begin{equation}}
\def\ee{\end{equation}}

\begin{document}

\begin{titlepage}
\renewcommand{\thefootnote}{\fnsymbol{footnote}}

\begin{center}
\textbf{\LARGE Hawking Radiation from General \\[0.5cm]Kerr-(anti)de Sitter Black
 Holes}\vspace{1cm}

 Zhibo Xu\footnote{Email:xuzhibo@pku.edu.cn}\hspace{0.5cm}and\hspace{0.5cm}
 Bin Chen\footnote{Email:bchen01@pku.edu.cn}\\
\vspace{0.5cm}
\emph{School of Physics, Peking University, Beijing 100871, P. R. China}\\[0.3cm]
\end{center}
\vspace{1.5cm} \centerline{\textbf{Abstract}}\vspace{0.5cm}
 We calculate the total flux of Hawking radiation from Kerr-(anti)de Sitter
 black holes by using gravitational anomaly method developed in \cite{Rob05}. We
 consider the general Kerr-(anti)de Sitter
 black holes in arbitrary $D$ dimensions with the maximal number ${[}D/2{]}$ of
 independent rotating parameters.
 We find that the physics near the horizon can be
 described by an infinite collection of $(1+1)$-dimensional quantum fields coupled
 to a set of gauge fields with
charges proportional to the azimuthal angular momentums $m_i$.
With
 the requirement of anomaly cancellation and regularity at the
 horizon, the Hawking radiation is determined.

\end{titlepage}
\setcounter{footnote}{0}

%%%%%%%%%%%%%%%%%%%%%%%%%%%%%%%%%%%%%%%%%%%%%%%%%%%%%%%%%%%%%%%%%%%%%%%%%%%%%%%%%%%%%%%%%%%%%%%%%%%%%%%%
%%%%%%%%%%%%%%%%%%%%%%%%%%%%%%%%%%%%%%%%%%%%%%%%%%%%%%%%%%%%%%%%%%%%%%%%%%%%%%%%%%%%%%%%%%%%%%%%%%%%%%%%

\section{Introduction}

 Hawking radiation is one of the most important and intriguing
 effects in black hole physics. It shows that black hole is
 not really black, it radiates thermally like
 black body. Precisely speaking, Hawking radiation is the quantum
 effect of field in a background space-time with a
future event horizon. It has a feature that the radiation is
determined universally by the horizon properties. It has several
derivations. The original one discovered by Hawking
\cite{Haw75,Haw74} is by directly calculating the Bogoliubov
coefficients between in and out states of fields in a black hole
background. This approach relies on the fact that in a curved
background the choice of vacuum for incoming and outgoing particle
is not unique. Later on a derivation based on the path-integral
quantum gravity was given in \cite{Har76}.  A few years ago, Parikh
and Wilczek\cite{Par00} proposed a tunneling picture in which
particle pair production happens near the horizon and Hawking
radiation could be obtained by calculating WKB amplitudes for
classically forbidden paths.

Very Recently, Robinson and Wilczek \cite{Rob05} have given a new
derivation of Hawking radiation in the Schwarzschild black hole
background through gravitational anomaly. This work is to some
extent inspired by the work of Christensen and Fulling
\cite{Chr77}, in which the radiation created in the
$(1+1)$-dimensional Schwarzschild black hole background was
determined by the trace anomaly and the energy-momentum
conservation law. In this approach, boundary conditions at the
horizon and the infinity are required to specify the Unruh
\cite{Unr76} vacuum. Moreover the method in \cite{Chr77} could not
be applied to the cases in more than $(2+1)$ dimensions. Robinson
and Wilczek found that by dimension reduction, the physics near
the horizon can be described by an infinite collections of free
$(1+1)$-dimensional fields because the mass and interaction terms
of quantum fields in the background are suppressed. If one only
consider the effective field theory outside the horizon, the
theory become chiral since classically all ingoing modes can not
affect physics outside the horizon. Quantum mechanically, the
effective theory is anomalous with respect to gauge or general
coordinate symmetries. The anomaly should be cancelled by the
quantum effects of the classically irrelevant incoming modes. The
condition for chiral and gravitational anomaly cancellation and
regularity requirement at the horizon, combining with the
energy-momentum conservation law, determines Hawking fluxes of the
charge and energy-momentum. Robinson and Wilczek's treatment once
again shows that  Hawking radiation (if we neglect the
back-reaction on the background) is universal, it only depends on
the property of the event horizon.

In the further development, Iso et al. \cite{Iso06,Iso06_1}
investigated the charged and rotating black hole. By using a
dimensional reduction technique, they found each partial wave of
quantum fields in $d=4$ rotating black hole background can be
interpreted as a $(1+1)$-dimensional charged field with a charge
proportional to the azimuthal angular momentum $m$. The total flux
of Hawking radiation can be determined by demanding gauge
invariance and general coordinate covariance at the quantum level.
And the boundary conditions are clarified. The results are
consistent with the effective action approach. Murata et
al.\cite{Mur06} extended the method to Myers-Perry black
holes\cite{Mye86} with only one rotating axis and also clarified
the boundary condition. The Hawking radiation from general
spherically symmetric black holes\cite{Vag06} and BTZ black holes
\cite{Set06} have also been investigated.

In this paper, we further extend Robinson and Wilczek's derivation
of Hawking radiation to general Kerr-(anti)de Sitter(K(A)dS) black
holes \cite{Gib04} in $D$ dimensions. For a general K(A)dS metric,
there are at most $N=[\frac{D-1}{2}]$ Killing symmetries
corresponding to the rotational symmetries in $N$ orthogonal
spatial 2-planes. For a quantum field in such backgrounds, the
physics near the future event horizon could still be effectively
described by an infinite collection of $(1+1)$-dimensional fields
coupled to $N$ $U(1)$ gauge fields. We discuss such dimensional
reduction in detail in  sec. \textbf{$2$}. In sec. \textbf {$3$},
we obtain the Hawking fluxes by requiring anomaly cancellation and
regularity condition. The final section is devoted to the
conclusion.

%%%%%%%%%%%%%%%%%%%%%%%%%%%%%%%%%%%%%%%%%%%%%%%%%%%%%%%%%%%%%%%%%%%%%%%%%%%%%%%%%%%%%%%%%%%%%%%%%%%%%%%%
%%%%%%%%%%%%%%%%%%%%%%%%%%%%%%%%%%%%%%%%%%%%%%%%%%%%%%%%%%%%%%%%%%%%%%%%%%%%%%%%%%%%%%%%%%%%%%%%%%%%%%%%

\section{Quantum Fields in general Kerr-(anti)de Sitter Black Holes}

In this section, we will discuss the quantum fields in general
Kerr-(anti)de Sitter blacks holes and its effective dimensional
reduction near the horizon. The Kerr-(anti)de Sitter metric in
D-dimension has been studied carefully in \cite{Gib04}. Here we
just give a brief review of its basic property. The metric takes
the form in an Boyer-Lindquist coordinates
\begin{eqnarray}
 ds^2 & = & -W(1-\lambda r^2)dt^2+\frac{2M}{VF}\left( Wdt-\sum_{i=1}^N \frac{a_i\mu_i^2 d\varphi_i}{1+\lambda a_i^2}
\right) ^2 +\sum_{i=1}^N \frac{r^2+a_i^2}{1+\lambda a_i^2}\mu_i^2 d\varphi_i^2 \nonumber \\
   & &{}+\frac{VF dr^2}{V-2M}+\sum_{i=1}^{N+\epsilon} \frac{r^2+a_i^2}{1+\lambda a_i^2}d\mu_i^2
   +\frac{\lambda}{W(1-\lambda r^2)}\left(\sum_{i=1}^{N+\epsilon} \frac{r^2+a_i^2}{1+\lambda a_i^2}
    \mu_i d\mu_i\right)^2 \label{kerr Metric}
\end{eqnarray}
where\\
 \[ \epsilon= \begin{cases} 1,\ D & \mbox{even} \\ 0,\ D & \mbox{odd} \end{cases} \]
 \[W\equiv \sum_{i=1}^{N+\epsilon}\frac{\mu_i^2}{1+\lambda a_i^2}\]
\[
   V\equiv r^{\epsilon -2}(1-\lambda r^2)\prod_{i=1}^N
  (r^2+a_i^2),\ \ \ \ F\equiv \frac{1}{1-\lambda
   r^2}\sum_{i=1}^{N+\epsilon}\frac{r^2\mu_i^2}{r^2+a_i^2}.
\]
Here $N$ is the integral part of $(D-1)/2$. There are $N$
independent rotation parameters $a_i$ in $N$ orthogonal spatial
2-planes. The $\varphi_i$'s are azimuthal angular coordinates. And
the $\mu_i$'s are the latitudinal coordinates obeying a constraint
$\displaystyle{\sum_{i=1}^{N+\epsilon}\mu_i^2=1}$, so only
$N+\varepsilon -1$ latitudinal coordinates $\mu_i$ are
independent. The $\lambda$ is the cosmological constant. Up to the
sign of $\lambda$, the above metric describes different Kerr black
holes in $D$ dimensions:
\[ \begin{cases} \lambda>0,\ \ \mbox{Kerr-de Sitter metric} \\
 \lambda=0,\ \ \mbox{Myers-Perry metric\cite{Mye86}} \\
 \lambda<0,\ \ \mbox{Kerr-Anti-de Sitter metric} \end{cases}\]
Hawking fluxes of Myers-Perry black holes with only one azimuthal
angular momentum has been discussed in \cite{Mur06}. In this
paper, we will discuss Hawking radiation of the other two cases
with any permissible angular momentums. Certainly our discussion
apply to the Myers-Perry black holes with more than one angular
momentums.

It is remarkable that for the black holes in de Sitter spacetime,
there exist a cosmological event horizon. However, our motivation
is to study the Hawking radiation of the black hole so we focus on
the physics near the black hole event horizon.

The metric (\ref{kerr Metric}) could be cast into a generalized
Boyer-Lindquist form,
\begin{eqnarray}
&&ds^2=Xdt^2+2Y_idtd\varphi^i+Z_{ij}d\varphi^i d\varphi^j+g_{ab}dx^a
dx^b+\frac{1}{B}dr^2 \label{Gen Metric}
\end{eqnarray}
where $\varphi^i, i=1,\cdots N$ are periodic with period $2\pi$
and $x^a, a=1,\cdots n$ with $n=N+\epsilon-1$ are independent
latitudinal coordinates. All the metric components only depend on
 $x^a$($\mu_i$) and the radial coordinate $r$. The
$Z_{ij}$ and $g_{ab}$ are positive definite and their corresponding
inverses are $Z^{ij}$ and $g^{ab}$ with $Z^{ij}Z_{jk}=\delta^i_k,\
g^{ab}g_{bc}=\delta_c^a$.

From the discussion in \cite{Gib04},  the angular velocities of
the horizon are given by $\Omega^i=\Omega_H^i$
\begin{eqnarray}
\Omega_H^i=-Y^i|_{r=r_H}=\frac{a_i(1-\lambda
r_H^2)}{r_H^2+a_i^2}\label{Angu Velocity}
\end{eqnarray}
where $Y^i=Z^{ij}Y_j$ and $r_H$ is the radius of the horizon.
Using (\ref{kerr Metric}), $r_H$ is just the largest positive root
of equation $V-2M=0$ or in the metric (\ref{Gen Metric}) the
largest positive root of equation $B=0$. The angular velocities,
relative to a non-rotating frame at infinity, is a little
different from (4.4) in \cite{Gib04}, which is defined relative to
a rotating frame at infinity. The null generator $l$ of the
horizon is a linear combination of the Killing vector fields
\begin{eqnarray}
l=\frac{\partial}{\partial
t}+\Omega_H^i\frac{\partial}{\partial\varphi^i}\label{Null
Generator}
\end{eqnarray}
The surface gravity on the horizon is
\begin{eqnarray}
\kappa^2=(\nabla^\mu L)(\nabla_\mu L)|_{r=r_H}\label{Surface gra}
\end{eqnarray}
where
\begin{eqnarray}
-L^2=l^\mu l_\mu=X+2Y_i\Omega_H^i+Z_{ij}\Omega^i_H\Omega_H^j
\label{L def}
\end{eqnarray}
Note that $L$ and $B$ vanish on the horizon but $\partial_rL$ and
$\partial_rB$ are non-zero. So near the horizon, we have
\begin{eqnarray}
L^2\approx(\partial_r L^2)|_{r=r_H}(r-r_H), \ \ \
B\approx(\partial_r B)|_{r=r_H}(r-r_H )\label{approx}
\end{eqnarray}
Thus the surface gravity is:
\begin{eqnarray}
\kappa=\frac{1}{2}\sqrt{(\partial_r L^2)(\partial_r
B)|_{r=r_H}}=\frac{1}{2}(1-\lambda
r_H^2)\frac{V'(r_H)}{V(r_H)}\label{Sur Gra}
\end{eqnarray}
The property that $\kappa$ is a constant and $Y^iY_i-X=0$ on the
horizon are very important to the following discussions.

 In order
to do dimensional reduction, we need some other properties of the
metric near the horizon. Define $A=Y^iY_i-X$ then
\begin{eqnarray}
L^2-A=-Z_{ij}(\Omega_H^i+Y^i)(\Omega_H^j+Y^j)
\end{eqnarray}
From the definition (\ref{Angu Velocity})
$\Omega_H^i+Y^i|_{r=r_H}=0$, so near the horizon
$\Omega_H^i+Y^i\approx C^i(r-r_H)$, then we have
\begin{eqnarray}
L^2-A\approx -Z_{ij}C^iC^j(r-r_H)^2,\ \ \ \ \partial_r
L^2=\partial_r A|_{r=r_H}\label{L^2=A}
\end{eqnarray}

When $A\neq0$, the inverse of the metric (\ref{Gen Metric}) can be
written as
\begin{eqnarray}
&& g^{tt}=-\frac{1}{A},\ \ \ \ \ g^{ij}=-\frac{1}{A}Y^iY^j+Z^{ij}
\nonumber\\ && g^{ti}=g^{i t}=\frac{1}{A}Y^i, \ g^{rr}=B,
\label{Inv Met}
\end{eqnarray}

Note that near the horizon $A\rightarrow 0$, $Y^i, Z^{ij}$ and
$g^{ab}$ are finite. This property is essential to the dimensional
reduction. The metric (\ref{Gen Metric}) can be written in another
form
\begin{eqnarray}
ds^2=-Adt^2+Z_{ij}(d\varphi^i+Y^idt)(d\varphi^j+Y^jdt)+\frac{1}{B}dr^2+g_{ab}dx^adx^b
\end{eqnarray}
with
\begin{eqnarray}
\sqrt{-g}=\sqrt{Ag_1g_2\frac{1}{B}}\label{Determinant}
\end{eqnarray}
where $g=\det(g_{\mu\nu}),\  g_1=\det(Z_{ij}),\  g_2=\det(g_{ab})$

Now let's consider a free complex scalar field for simplicity in the
general Kerr-(anti)de Sitter black hole background. Using the
inverse of metric (\ref{Inv Met}), the free part of the action is
\begin{eqnarray}
S_{free}&=&-\int
d^Dx\sqrt{-g}g^{\mu\nu}\partial_\mu\phi^*\partial_\nu\phi
\nonumber\\
 &=&-\int dtdrd^N\varphi^id^nx^a\sqrt{-g}\left(-\frac{1}{A}\partial_t\phi^*\partial_t\phi+\frac{1}{A}Y^i\partial_t
 \phi\partial_i\phi^*+\frac{1}{A}Y^i
 \partial_{i}\phi\partial_t\phi^*\right.\nonumber\\
 &&\left.
 -\frac{1}{A}Y^iY^j\partial_{i}\phi^*\partial_{j}\phi+Z^{ij}\partial_{i}\phi^*
 \partial_{j}\phi+B\partial_r\phi^*\partial_r\phi+g^{ab}\partial_{a}\phi^*\partial_{b}\phi\right)
\end{eqnarray}
where $\partial_i$ denotes $\frac{\partial}{\partial \varphi^i}$
and $\partial_a$ denotes $\frac{\partial}{\partial x^a}$.

In order to consider the physics near the horizon, we make a
coordinate transformation $\frac{dr_*}{dr}=f(r)^{-1}$, where
$f(r)\equiv\sqrt{A'B'}|_{r=r_H}(r-r_H)=2\kappa (r-r_H)$. In this
frame, considering the region near the outer horizon $r_H$, the
finite terms $Z^{ij}\partial_{i}\phi^*
 \partial_{j}\phi$ and
 $g^{ab}\partial_{a}\phi^*\partial_{b}\phi$
 are suppressed by the factor $f(r(r_*))$, vanishing exponentially
 fast near the horizon.  We can also substitute $\sqrt{AB},\
 g_1g_2$ by $f(r),\ g_1g_2|_{r=r_H}$ because the omitting terms are
 suppressed for the same reason. Similarly, one can redefine $Y^i$ by $-\frac{a_i(1-\lambda
 r^2)}{r^2+a_i^2}$.
 So the action with dominant terms is
\begin{eqnarray}
S&=&-\int
dtdrd^N\varphi^id^nx^a\sqrt{g_1(r_H)g_2(r_H)}\left[-f(r)^{-1}\left(\partial_t-Y^i\partial_{i}\right)\phi^*
\left(\partial_t-Y^i\partial_{i}\right)\phi\right.\nonumber\\
&&\left.+f(r)\partial_r\phi^*\partial_r\phi\right]
\end{eqnarray}

 We can expand
$\phi$ by a complete set of orthogonal functions of $(\varphi^i,\
x^a)$ with the measure $\sqrt{g_1(r_H)g_2(r_H)}$. As we know, the
angles $\varphi^i$ are periodic with period $2\pi$ and coordinates
$x^a$ come from $\mu_i$ which obey a constraint $\sum_i\mu_i^2=1$.
So $(\varphi^i,x^a)$ describe a compact manifold with a metric
\begin{eqnarray}
ds^2=Z_{ij}(r_H)d\varphi^i d\varphi^j+g_{ab}(r_H)dx^a dx^b
\label{sur met}
\end{eqnarray}
whose measure is $\sqrt{g_1(r_H)g_2(r_H)}$. Then the
eigenfunctions of the operator $\nabla^2$ of the compact manifold
with the metric (\ref{sur met}) comprise a complete orthogonal
functions. Note that there are N killing vectors
$\frac{\partial}{\partial{\varphi^i}}$ which generate isometry.
The eigenfunctions can be given by
\begin{eqnarray}
&&Y_{m_1\cdots
m_N\alpha}=\prod_{j=1}^N\exp^{im_j\varphi^j}f_\alpha
(x^a)\end{eqnarray} satisfying
 \begin{equation} \int
dx^ad\varphi^i\sqrt{g_1(r_H)g_2(r_H)}Y^*_{m_1\cdots
m_N\alpha}Y_{n_1\cdots n_N\beta}=\delta_{m_1n_1}\cdots
\delta_{m_Nn_N}\delta_{\alpha\beta}
\end{equation}
Performing the partial wave decomposition of $\phi$ in terms of
these functions,
\begin{eqnarray}
\phi=\sum_{m_1,\cdots, m_N,\alpha}\phi_{m_1\cdots
m_N\alpha}Y_{m_1\cdots m_N\alpha},
\end{eqnarray}
the theory is reduced to a two-dimensional effective theory with
an infinite collection of fields with quantum numbers$(m_1,\cdots,
m_N,\alpha)$, simply denoted as $\phi_n$. It is straightforward to
show that the physics near the outer horizon can be effectively
described by an infinite collection of massless
$(1+1)$-dimensional fields with the following action
\begin{eqnarray}
S=-\int dtdr \left[
-f(r)^{-1}\left(\partial_t-im_jY^j\right)^*\phi_n^*
\left(\partial_t-im_jY^j\right)\phi_n+f(r)\partial_r\phi_n^*\partial_r\phi_n\right]
\label{Action}
\end{eqnarray}

%%%%%%%%%%%%%%%%%%%%%%%%%%%%%%%%%%%%%%%%%%%%%%%%%%%%%%%%%%%%%%%%%%%%%%%%%%%%%%%%%%%%%%%%%%%%%%%%%%%%%%%%
%%%%%%%%%%%%%%%%%%%%%%%%%%%%%%%%%%%%%%%%%%%%%%%%%%%%%%%%%%%%%%%%%%%%%%%%%%%%%%%%%%%%%%%%%%%%%%%%%%%%%%%%

\section{Anomalies and Hawking fluxes}

In this section, we will try to obtain the Hawking fluxes. We will
follow the approach firstly proposed in \cite{Iso06,Iso06_1}. The
basic point is that  the Hawking fluxes can be determined by the
anomaly cancellation of the effective chiral theory.

From the effective action(\ref{Action}), near the horizon, each
partial wave mode of the scalar field $\phi_n$ can be considered
as $(1+1)$-dimensional complex scalar field in the backgrounds of
the metric $g_{\mu\nu}$ and $N$ gauge potentials $A^i_{\mu}$
\begin{eqnarray}
&&g_{tt}=-f(r),\ \ g_{rr}=f(r)^{-1},\ \ g_{rt}=0 \nonumber\\
&&A^i_t=Y^i,\ \ \ \ \ \ \ \ A^i_r=0
\end{eqnarray}
In this case, there are $N\ \ U(1)$ gauge symmetries and $N$ gauge
currents which actually relate to angular momentum currents. Each
gauge symmetry originates from the axial isometry along
$\varphi^i$ direction. With respect to gauge fields $A^i_{\mu}$,
the field $\phi_n$ has charges $m_i$, which is the azimuthal
quantum number rotating along $\varphi^i$ direction. The
corresponding $U(1)$ currents $J^r_i$  can be defined from the
$D$-dimensional energy-momentum tensor.
\begin{eqnarray}
J_i^r=-\int d^nx^ad^N\varphi^i\sqrt{-g}T^r_{\varphi^i}.
\end{eqnarray}
In effect, performing a partial wave decomposition and an
integral, we find the result of right side of the above equation
is just the current obtained from the two-dimensional effective
action. Similarly the energy-momentum tensor in two-dimensional
effective action is the reduction of the one in $D$-dimension
\begin{eqnarray}
T_{t(2)}^r=\int d^nx^ad^N\varphi^i\sqrt{-g}T^r_t.
\end{eqnarray}
 Without bringing any
confusion, from now on we denote $T^r_{t(2)}$ as $T_t^r$ for
simplicity.

As shown in \cite{Iso06}, we can divide the region
$r\in[r_H,\infty]$ into two regions. One is
$r\in[r_H+\varepsilon,\infty]$ which is apart from the horizon and
the other is $r\in[r_H,r_H+\varepsilon]$ which is near the
horizon. In the region $r\in[r_H+\varepsilon,\infty]$, each
current is conserved. So we have
\begin{eqnarray}
\partial_rJ^r_{i(o)}=0\label{conserve equ1}
\end{eqnarray}
On the other hand, in the near horizon region, the effective two
dimensional theory become chiral since classically the ingoing modes
are irrelevant and there are only outgoing modes. In this effective
chiral theory, the gauge symmetries and general coordinate
transformation symmetries become anomalous quantum mechanically. The
anomaly equation for each $U(1)$ current near the horizon
is\cite{Ber00,Fuj}
\begin{eqnarray}
\partial_rJ^r_{i(H)}=\frac{m_i}{4\pi}\partial_r\mathcal {A}_t \label{anomaly
equ2}
\end{eqnarray}
where $\mathcal {A}_t=m_iA_t^i$ is the sum of $N$ $U(1)$'s.

Actually one can take $\mathcal {A}(r)$ more seriously. From
effective action, we can take the point of view that the scalar
field coupled to a single gauge potential $\mathcal {A}(r)$. There
exist a $U(1)$ gauge symmetry associated with gauge potential
$\mathcal {A}(r)$. The corresponding current denoted as $\mathcal
{J}(r)$ can be constructed from the original $N\ \ U(1)$ gauge
symmetries. Note that each $J^r_i$ is not independent for a fixed
azimuthal angular momentum $m_i$. Their expectation values are
related as $\frac{1}{m_i}J^r_i=\frac{1}{m_j}J^r_j=\mathcal {J}^r$.
The anomaly equation for $\mathcal {J}(r)$ is \be\partial_r\mathcal
{J}^r=\frac{1}{4\pi}\partial_r\mathcal {A}_t,\ee so we have anomaly
equation (\ref{anomaly equ2}). Solving the above equations in each
region, we have,
\begin{eqnarray}
J^r_{i(o)}&=&C_{i(o)} \nonumber \\
J_{i(H)}^r&=&C_{i(H)}+\frac{m_i}{4\pi}\left(\mathcal
{A}_t(r)-\mathcal {A}_t(r_H)\right)\label{current}
\end{eqnarray}
where $C_{i(o)}$ and $C_{i(H)}$ are two integration constants.
$C_{i(H)}$ is the value of the consistent current of the outgoing
modes at the horizon and $C_{i(o)}$ is the value of the angular
momentum flux at infinity. It is our goal to determine $C_{i(o)}$,
which encodes the information of Hawking radiation.

Under gauge transformations, the variation of the effective action
is given by
\begin{eqnarray}
-\delta W=\int \sqrt{-g_{(2)}}\lambda \nabla_\mu J^\mu_i
\end{eqnarray}
where $\lambda$ is a gauge parameter, and \be
J^\mu_i=J^\mu_{i(o)}\theta_+(r)+J^\mu_{i(H)}H(r).\ee Here
$\theta_+(r)=\theta (r-r_H-\varepsilon)$ and $H(r)=1-\theta_+(r)$,
where $\theta(x)$ is a step function. Note that we have not take
the contribution of the ingoing modes into account.  Using
equations (\ref{conserve equ1})(\ref{anomaly equ2}) we have
\begin{eqnarray}
-\delta W=\int d^2x\lambda \left[\delta
(r-r_H-\epsilon)\left(J^\mu_{i(o)}-J^\mu_{i(H)}+\frac{m_i}{4\pi}\mathcal
{A}_t\right)+\partial_r\left(\frac{m_i}{4\pi}\mathcal
{A}_tH(r)\right)\right]
\end{eqnarray}
Since the underlying theory must be gauge invariant, so $\delta
W=0$. Actually the last term is cancelled by quantum effects of the
classically irrelevant ingoing modes \cite{Rob05}. Then the
coefficient of the delta-function should vanish. With the results
(\ref{current}), we can obtain a relation between the two constants
\begin{eqnarray}
C_{i(o)}=C_{i(H)}-\frac{m_i}{4\pi}\mathcal {A}_t(r_H) \label{flux}
\end{eqnarray}

In order to determine the value of $C_{i(o)}$, one need to impose
the regularity condition. As discussed in \cite{Iso06,Iso06_1}, the
regularity requires that the covariant current is zero on the
horizon,
\begin{eqnarray}
\widetilde{J^r_i}=J^r_i+\frac{m_i}{4\pi}\mathcal {A}_tH(r),\ \
\widetilde{J^r_i}(r_H)=0
\end{eqnarray}
Then the flux of the angular momentum is obtained as
\begin{eqnarray}
C_{i(o)}=-\frac{m_i}{2\pi}\mathcal
{A}_t(r_H)=\frac{m_i}{2\pi}\sum_{j=1}^N m_j\frac{a_j(1-\lambda
r_H^2)}{r_H^2+a_j^2}
\end{eqnarray}

Similarly we can determine the flux of the energy-momentum tensor
radiated from general Kerr-(anti)de Sitter black holes. In the
presence of the effective gauge potentials $A^i_t(r)$, the
conservation equation outside the horizon is modified to be
\begin{eqnarray}
\partial_rT^r_{t(o)}=\mathcal {F}_{rt}\mathcal {J}^r_{(o)}
\end{eqnarray}
where $\mathcal {F}_{rt}=\partial_r\mathcal {A}_t$. Note that the
right hand side of the above relation depends simply on $\mathcal
{A}_t$. With the definition of $\mathcal {J}^r$, we have $\mathcal
{J}^r_{(o)}=-\frac{1}{2\pi}\mathcal {A}_t(r_H)\equiv C_o$, where \be
C_o= \sum_{j=1}^N m_j\frac{a_j(1-\lambda r_H^2)}{r_H^2+a_j^2}. \ee
The solution of the above equation gives the value of the energy
flux at spatial infinity
\begin{eqnarray}
T_{t(o)}^r=a_o+C_o\mathcal {A}_t(r)
\end{eqnarray}
where $a_o$ is an integration constant. Physically, it could be
taken as the value of the total energy flow of radiation measured
at spatial infinity.

On the other hand, there are gauge and gravitational anomalies near
the horizon and the anomaly equation is now as
\begin{eqnarray}
\partial_rT^r_{t(H)}=\mathcal {F}_{rt}\mathcal {J}^r_{(H)}+\mathcal
{A}_t\nabla_\mu\mathcal {J}^\mu_{(H)}+\partial_r N^r_t
\end{eqnarray}
where $N^r_t=(f'^2+ff'')/192\pi$\cite{Iso06}. The second term
indicates gauge anomaly while the third term is gravitational
anomaly\cite{Alv} for the consistent energy-momentum tensor. From
the definition of $\mathcal {J}^r$ and Eqs.
(\ref{current})(\ref{flux}) we have $\mathcal
{J}^r_{(H)}=C_o+\frac{1}{4\pi}\mathcal {A}_t(r)$. $T^r_{t(H)}$ can
be solved as
\begin{eqnarray}
T^r_{t(H)}=a_H+\int_{r_H}^rdr\partial_r\left(C_o\mathcal
{A}_t+\frac{1}{2\pi}\mathcal {A}_t^2+N^r_t\right) \label{energy}
\end{eqnarray}
where $a_H$ is an integration constant.

Under the general coordinate transformation, the variation of the
effective action is
\begin{eqnarray}
-\delta W&=&\int d^2x\sqrt{-g_{(2)}}\xi^t\nabla_\mu
T^{\mu}_t\nonumber \\
&=&\int d^2x \xi^t\left[C_0\partial_r\mathcal
{A}_t(r)+\partial_r\left(\frac{1}{4\pi}\mathcal
{A}_t^2H(r)+N^r_tH(r)\right)\right. \nonumber\\
&&\left.\ \ \ \ +\left(T^r_{t(o)}-T^r_{t(H)}+\frac{1}{4\pi}\mathcal
{A}_t^2+N^r_t\right)\delta(r-r_H-\epsilon)\right]
\end{eqnarray}
where $\xi^t$ is the transformation parameter and
$T_\nu^\mu=T_{\nu(o)}^\mu \theta_+(r)+T_{\nu(H)}^\mu H(r)$. The
first term is generated by classical current. The second term should
be cancelled by the quantum effect of the ingoing modes. As we
discussed before, the last term should vanish because the underlying
theory is general coordinate transformation covariant. So we have:
\begin{eqnarray}
a_o=a_H+\frac{1}{4\pi}\mathcal {A}_t^2(r_H)-N^r_t(r_H)
\end{eqnarray}

Similarly we need to impose the regularity condition, which
requires that the covariant energy-momentum flux vanish on the
future horizon. The covariant energy-momentum is defined by
\cite{Bar,Ber01}
\begin{eqnarray}
\widetilde{T^r_t}=T^r_t+\frac{1}{192\pi}(ff''-2f'^2)
\end{eqnarray}
Combining with (\ref{energy}), the condition reads
\begin{eqnarray}
a_H=\kappa^2/24\pi=2N^r_t(r_H),
\end{eqnarray}
where $\kappa=2\pi/\beta$ is the surface gravity of the black hole.
Therefore the total flux of the energy-momentum tensor is given by
\begin{eqnarray}
a_o=\frac{\mathcal
{A}^2_t(r_H)}{4\pi}+N^r_t(r_H)=\frac{1}{4\pi}\left(\sum_{i=1}^N
m_i\Omega^i_H\right)^2+ \frac{\pi}{12\beta^2}
\end{eqnarray}

%%%%%%%%%%%%%%%%%%%%%%%%%%%%%%%%%%%%%%%%%%%%%%%%%%%%%%%%%%%%%%%%%%%%%%%%%%%%%%%%%%%%%%%%%%%%%%%%%%%%%%%%
%%%%%%%%%%%%%%%%%%%%%%%%%%%%%%%%%%%%%%%%%%%%%%%%%%%%%%%%%%%%%%%%%%%%%%%%%%%%%%%%%%%%%%%%%%%%%%%%%%%%%%%%

\section{Conclusion}

In this paper we studied the Hawking radiation of general
Kerr-(anti)de Sitter black holes. We considered a complex scalar
field in a general Kerr-(anti)de Sitter black hole background.
Near the horizon, the field can be described by an infinite
collection of $(1+1)$-dimensional fields. A formal derivation of
the dimensional reduction based on the properties of the horizon
is given and each term in the effective action has obvious
physical meaning. Although there are no gauge fields in the
original $D$ dimensional theory, the metric has $N$ isometries,
which induce $N$ $U(1)$ gauge symmetries in the effective
two-dimensional theory. Each partial wave mode is charged under
the gauge symmetries with charge $m_i$'s, where the $m_i$'s are
angular quantum numbers. We have shown that Hawking radiation from
general Kerr-de Sitter space can be determined by the cancellation
condition of the gravitational anomaly and gauge anomaly,
combining with the boundary condition required by regularity at
the horizon. We obtained the Hawking flux of each angular momentum
$C_{i(o)}$ and energy-momentum tensor $a_o$ for each partial wave
mode:
\begin{eqnarray}
C_{i(o)}&=&\frac{m_i}{2\pi}\sum_{j=1}^N m_j\frac{a_j(1-\lambda
r_H^2)}{r_H^2+a_j^2} \\
a_o&=&=\frac{1}{4\pi}\left(\sum_{i=1}^N m_i\Omega^i_H\right)^2+
\frac{\pi}{12\beta^2}.
\end{eqnarray}

Our work shows that the proposal in \cite{Rob05,Iso06,Iso06_1} can
be applied to more general blackhole backgrounds in higher
dimensions which have more than one angular momentum. It would be
interesting to apply the gravitational anomaly method to study
other problems in blackhole physics.

\section*{Acknowledgements}

 The work was partially supported by NSFC Grant No.
10405028,10535060, NKBRPC (No. 2006CB805905) and the Key Grant
Project of Chinese Ministry of Education (NO. 305001).

\end{document}